\documentclass[conference]{IEEEtran}

\usepackage{amsmath,graphicx}

\usepackage{amssymb}
\usepackage{amsmath}
\usepackage{graphicx}
\usepackage{multirow}
\usepackage{float}
\usepackage{subfig}
\usepackage{array}
\usepackage{setspace}
\usepackage{float}
\restylefloat{table}
\ifCLASSINFOpdf
\else
\fi
\begin{document}
\title{Exploring Deep Learning Image Super-Resolution for Iris Recognition}

\author{\IEEEauthorblockN{Eduardo Ribeiro$^{1,2}$, Andreas Uhl$^{1}$, Fernando Alonso-Fernandez$^{3}$, Reuben A. Farrugia$^{4}$}
	\IEEEauthorblockA{$^{1}$University of Salzburg - Department of Computer Sciences -
		Salzburg, Austria\\
		$^{2}$ Federal University of Tocantins - Department of Computer Sciences - Tocantins, Brasil \\$^{3}$ Halmstad University - Halmstad, Sweden \\$^{4}$ University of Malta - Department of CCE - Msida, Malta \\}
}


%Telephone: (800) 555--1212, Fax: (888) 555--1212}


\maketitle

\begin{abstract}
In this work we test the ability of deep learning methods to provide an end-to-end mapping between low and high resolution images applying it to the iris recognition problem. Here, we propose the use of two deep learning single-image super-resolution approaches: Stacked Auto-Encoders (SAE) and Convolutional Neural Networks (CNN) with the most possible lightweight structure to achieve fast speed, preserve local information and  reduce artifacts at the same time.
We validate the methods with a database of 1.872 near-infrared iris images with quality assessment and recognition experiments showing the superiority of deep learning approaches over the compared algorithms.
\end{abstract}


% \fi
\IEEEpeerreviewmaketitle

\section{Introduction}
\label{sec:intro}

 Iris recognition technology is considered one of the most accurate and reliable biometric modalities for authentication today mainly due its stability and high degree of freedom in texture  \cite{Nguyen} \cite{Bowyer}.
Currently, most systems require the user to present their iris for the sensor at a close distance, however currently there is a constant pressure to make that relaxed conditions of acquisitions in such systems could be allowed \cite{Alonso}. One of the major problem in these conditions (for example at distance or on the move) is related to the quality of the images which are degraded as well as their resolutions which became low, i.e. the number of pixels in the iris region to allow a good recognition rate is constantly degraded when the resolution decreases as shown in \cite{Nguyen}.

Currently, several methods have been proposed for example based single-image super-resolution using different approaches as internal patch recurrence \cite{Ahuja}, regression functions \cite{Li2015284} \cite{Timofte} and sparse dictionary methods \cite{JYang}. The application of SR techniques to biometric systems is limited, with most research concentrated on faces \cite{Wang2014}. In the case of iris, some approaches exist  \cite{Nguyen2012} but they use whole images for reconstruction. Recently, a method based on PCA eigen transformation of local patches was proposed \cite{Alonso}, where each patch is reconstructed separately, providing better quality and detail, and lower distortions.

The first studies applying deep learning related to super-resolution in general were performed for image restoration. For example, fully-connected multilayer perceptrons were used for image denoising \cite{Burger} and Convolutional Neural Networks (CNN) were applied for natural image denoising \cite{Viren}.

Also, Stacked Auto-Encoders (SAE) were used for example-based super-resolution as can be seen in \cite{Fleet}, where in each layer a non-local self-similarity search with a collaborative local autoencoder is used to suppress the noise and enhance high-frequency texture details of patches.

Robust methods using deep-learning were also implemented to map a model from Low Resolution to High Resolution  patches trying to find the best regression functions to this mapping as in \cite{Jiwon}, \cite{Justin}, \cite{Christian}, \cite{Wenzhe}. Among these several successful examples, the Super-Resolution Convolutional Neural Network (SRCNN) \cite{Dong} has proved to be a good alternative for an end-to-end approach in super-resolution. 

In this work, we explore two typical deep learning approaches: Stacked Auto-encoders and Convolutional Neural Networks to increase the resolution and quality of low-resolution images by simulating long distance acquisition sensors. We use the CASIA-IrisV3-Interval database \cite{Casia} of NIR images for our experiments to validate the methods. Tests performed both in relation to the quality of the images as well as the iris recognition accuracy were carried out to see if the  performance is not degraded significantly in high upscaling factors.

\section{Methodology}
\label{sec:Methodology}

The single-image super-resolution methods presented in this paper aim at generating a High Resolution image (HR) from one low resolution input (LR). For this purpose, the image is upscaled using bicubic interpolation to the desired factor, then this image will pass through the deep learning (CNN or SAE) procedure that will try to correct the imperfections and noising to reconstruct the final super-resolution image. To do this reconstruction it is necessary to learn a mapping function $F$ where, given a LR image $Y$ (upscaled by bicubic interpolation), the goal of the method is to transform $Y$ into an image $F(Y)$ that is the closest possible to the ground truth HR image $X$.

For the evaluation of the methods in the CASIA-IrisV3-Interval database, first the images were downscaled through bicubic interpolation for the factors 2 (115x115), 4 (57x57), 8 (29x29) and 16 (15x15) and then re-upscaled through bicubic interpolation to the original size (231x231) to pass trough the deep learning procedure. If the CNN and SAE are trained only with factor 2, to achieve greater factors, the input images have to pass trough the network $log_2(n)$ times to achieve the desired factor $n$. For example, in a CNN trained with factor 2, to achieve the factor 8, the input image will first pass trough the CNN in order to achieve the factor 2, then the resultant image will pass again to the CNN to achieve the factor 4 and so on.

In this work we take advantage of a common strategy  used in image restoration, which is the extraction of patches and their representations as a series of pre-trained bases (such as PCA, DCT, Haar among other). Such filters are convolved with the image and in the case of this work will be optimized so that the mapping is the best possible. This can be done in one, two, or more layers and in the case of this work are followed by a reconstruction step which the predicted overlapping high-resolution patches are averaged to produce the final image.  This strategy is used both in the SAE's and CNN's that will be explained in the next subsections.

\subsection{Convolutional Neural Networks}

CNN’s are formed basically of a series of convolutional layers in the first levels (usually with a subsampling step) followed by one or more fully-connected neural networks similar to the multilayer neural networks \cite{lecun}.

\begin{figure}[ht]
	\centering
	\includegraphics[scale=0.99]{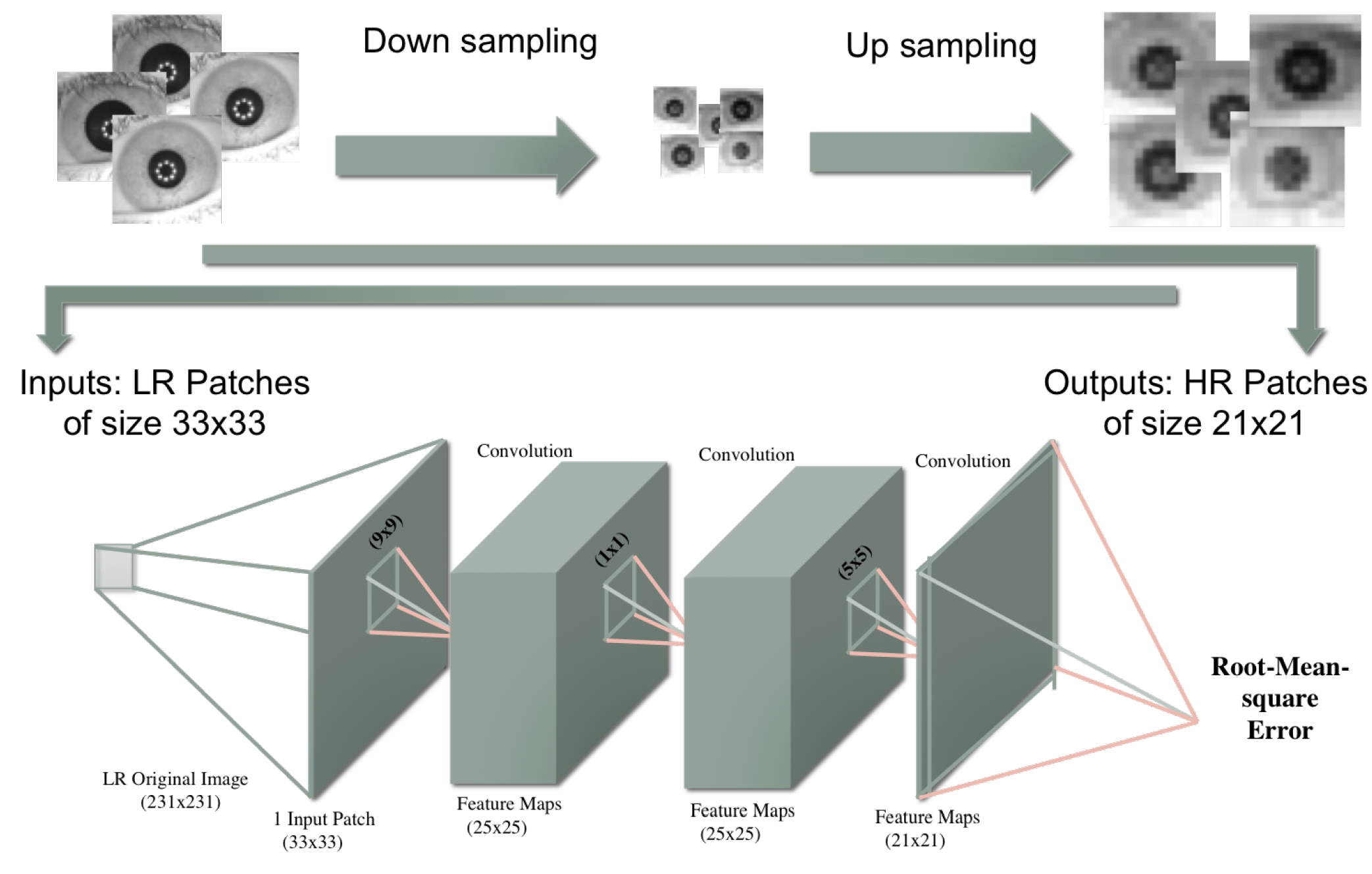}
	\caption{An illustration of the Convolutional Neural Network architecture for Iris Super-Resolution.}
	\label{cnn}
\end{figure}

The input of a CNN is a ($m \times m\times d$) patch where $(m \times m)$ is the dimension of the patch and $d$ the number of channels (depth) of the image. In this work, for the CNN training, patches are extracted from the HR images where $m=33$ and $d=1$, then the patches are downscaled (depending on the factor chosen for the method) and re-upscaled to the original size both using bicubic interpolation as it can be seen in the Figure \ref{cnn}.

%A series of $k$ filters of size ($n \times n\times e$) is applied to the inputs in a process called convolution that perform the product between the inputs and each filter resulting in a new output called activation map or feature map. The dimension of the feature map will depend on the stride and padding chosen in the convolution process. 

In this work, the implemented CNN has three convolutional layers, where: the first layer consists of 64 filters of size 9x9x1 with stride 1 and padding 0, the second layer with 32 filters of size 1x1x64 with stride 1 and padding 0, and the last layer with 1 filter of size 5x5x32 with stride 1 and padding 0. With all paddings set to zero, the feature maps will decrease in size resulting in a patch of size 21x21.  In the test phase, the overlapping patches will be extracted with stride 1 and only the central pixel of the resulting feature map will be used which means that the smaller size of the result feature map will not influence the final result image.

After each convolutional layer a non-linearity (or activation) function is applied to the feature maps mainly to accelerate the convergence of the stochastic gradient algorithm called ReLU rectifier function: $f(x) = \max (0,x)$, where $x$ is the neuron input. 

%One of the most used activation functions in the CNN's is the ReLU rectifier function $f(x) = \max (0,x)$ where $x$ is the neuron input that is demonstrably more efficient than other activation functions \cite{Sainath}. 

For the training with the high-resolution patches with their correspondent low-resolution patches we use the Mean Squared Error (MSE) as the loss function trying to achieved the best PSNR as possible when the CNN is completely trained and the loss minimization is done using stochastic gradient descent with the standard backpropagation method.

In this work we tested three different approaches for the CNN training:

\begin{itemize}
	\item From scratch (CNN FS): When the CNN weights are initialized randomly and trained according to the target image database (in the case of this work: the CASIA Interval V3 Iris Database) for the kernels domain adaptation, that is, to find the best way to map the data in order to perform the super-resolution.
	\item Transfer Learning (CNN TL): When an \textbf{off-the-shelf CNN} is chosen, which means that a CNN was pre-trained with a different database (in the case of this work: the ImageNet Database \cite{NIPS2012}) is used to perform the super-resolution in the target image database. 
	\item Fine Tuning (CNN FT): The pre-trained network (off-the-shelf CNN) training is continued with new entries (with the target image database) for the weights to adjust properly to the new scenario reinforcing the more generic features with a lower probability of overfitting.
	
\end{itemize}

\subsection{Stacked Auto-Enconders}

An auto-encoder is, by definition, a simple Neural Network (NN) designed to rebuild its own input in its output layer. For this reason, the number of neurons in the input layer is always the same as the output layer.  Successful applications demonstrate that Stacked Auto-Encoders can be a powerful alternative to deep learning \cite{Bengio2014}. 

%In autoencoders, if the encoding (first) and decoding (last) layers are transposed (the weight matrix is constrained), then the auto-encoder is said to have tied weights. Successful applications demonstrate that Stacked Auto-Encoders can be a powerful alternative to deep learning \cite{Bengio2014}. Stacking auto-encoders together gives the final model the advantage of extracting features in several hierarchical levels with the low level features being represented in the first layers and the high level features being represented in the subsequent layers. The auto-encoder training is performed in two stages: a layer-wise unsupervised training followed by a supervised discriminative fine-tuning back-propagation step \cite{Kandaswamy}.
%

%The Layer-wise pre-training of Stacked Auto-Encoders consists of several steps. First, the bottommost self-encoder is trained with the inputs. After self-encoding, a new self-encoder is implemented using the output from the decoder layer by taking the latent representation of the previous auto-encoder as input. The next auto-encoder is trained as before and the above steps are repeated until enough layers are pre-trained. The bottom of Figure \ref{SAE} shows an example of these steps. It should be noted that by combining the self-encoders, the configuration will be like a multilayer neural network that can be fully trained (fine-tuning) using a supervised backpropagation algorithm to achieve the desired results.

\begin{figure}[ht]
	\centering
	\includegraphics[scale=0.99]{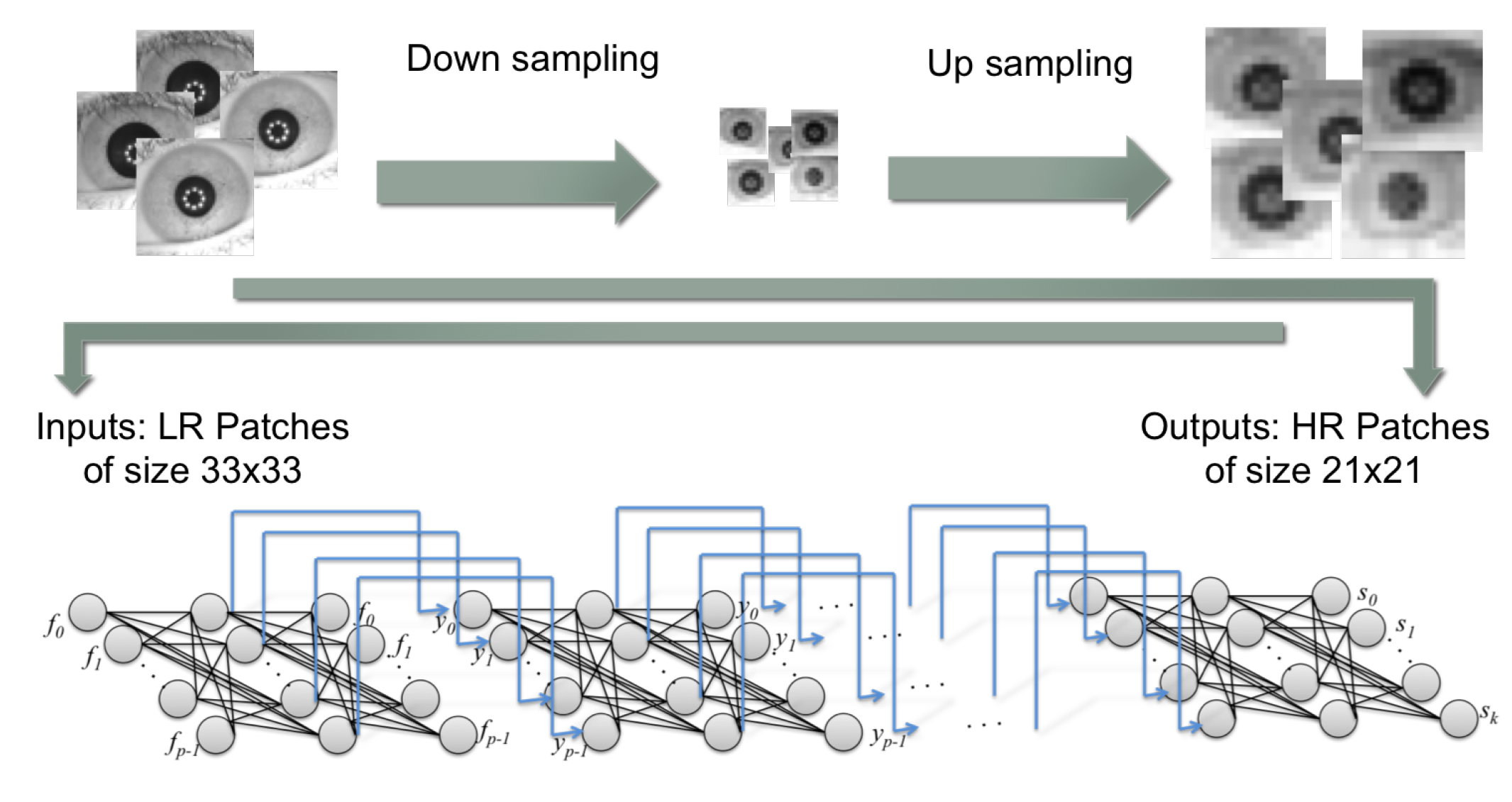}
	\caption{An illustration of the Stacked Auto-Encoder architecture for Iris Super-Resolution.}
	\label{SAE}
\end{figure}

For the Layer-wise pre-training of Stacked Auto-Encoders we use the HR patches downscaled and upscaled again using bicubic interpolation in the same way as for the CNN, however in this case, the matrix is turned to a vector in order to fit in the auto-encoder architecture. These vectors are used for the first auto-encoder as can be seen in Figure \ref{SAE} that are trained until a threshold is reached. In the second auto-encoder, we use the vector that we got from the hidden layer of the previous trained auto-encoder as input, and proceed in the first auto-encoder. The same process is applied to the third layer and so on. Then we use the original images (HR patches) as the targets in the last layer of the output auto-encoder. These targets are used to update the parameter of the deep multi-layered neural network (Stacked Auto-Encoders) by means of a supervised error backpropagation algorithm. This process tries to reconstruct the image patch by generalizing the missing pixels with the auto-encoder weights learned from the all images of the training database. 

When the training is completed, the auto-encoder is used to propagate all the LR patches upscaled using bicubic intepolation resulting in the reconstructed super-resolution patches in a magnification of 2 (when the training is done with this magnification). To achieve a magnification factor of 4, it is necessary to reinsert the reconstructed super-resolution images to the network in the same way as explained for the CNN approach.

For the experiments we trained four auto-encoders with the empirically chosen configuration: 1089-1000-1089 (where 1089 means the 33x33 input patches), 1000-2000-1000, 2000-2600-2000, 2600-2000-2600.  Consequently, in the fine-tuning phase, the NN configuration  for the Stacked Auto-Encoder experiment is: 1089-1000-2000-2600-2000-441. (21x21 output patches).  For the training we use the gradient descendent and backpropagation algorithms with the learning rate set to 0.2 and the number of training epochs set to 150 for each experiment.

\section{Experimental Setup}
\label{sec:Experimental Setup}

\begin{table*}[]
	\centering
	\caption{Results with different downscaling factors and two different factors (average values on the test dataset).}
	\label{qualityA}
	\scalebox{0.7}{
		\begin{tabular}{|c|c|c|c|c|c|c|c|c|c|c|c|c|c|c|c|c|}
			\hline
			\begin{tabular}[c]{@{}c@{}}LR Size\\ (scaling)\end{tabular}              &      & Bilinear       & Bicubic         & PCA            & SAE   & \begin{tabular}[c]{@{}c@{}}CNN FS \\ Factor 2\end{tabular} & \begin{tabular}[c]{@{}c@{}}CNN FS\\ Factor 4\end{tabular} & \begin{tabular}[c]{@{}c@{}}CNN TL\\ Factor 2\end{tabular} & \begin{tabular}[c]{@{}c@{}}CNN TL\\ Factor 4\end{tabular} & \begin{tabular}[c]{@{}c@{}}CNN FT\\ Factor 2\end{tabular} & \begin{tabular}[c]{@{}c@{}}CNN FT\\ Factor 4\end{tabular}  & \begin{tabular}[c]{@{}c@{}}CNN FS\\ Factor 8\end{tabular}   & \begin{tabular}[c]{@{}c@{}}CNN FT\\ Factor 8\end{tabular}  & \begin{tabular}[c]{@{}c@{}}CNN FS\\ Factor 16\end{tabular}  & \begin{tabular}[c]{@{}c@{}}CNN FT\\ Factor 16\end{tabular} \\ \hline
			\multirow{3}{*}{\begin{tabular}[c]{@{}c@{}}115x115\\ (1/2)\end{tabular}} & psnr & 32.17          & 34.04           & 34.63          & 32.56 & 35.51                                                      & -                                                         & 35.63                                                     & -                                                         & \textbf{35.93}                                            & -                        &    -                & - & - & -               \\ \cline{2-16}  
			& ssim & 0.892          & 0.926           & 0.934          & 0.897 & 0.945                                                      & -                                                         & 0.946                                                     & -                                                         & \textbf{0.948}                                            & -                                                 &   -      & - & - & -  \\ \cline{2-16} 
			& vif  & 0.813          & 0.819           & 0.771          & 0.724 & 0.821                                                      & -                                                         & 0.823                                                     & -                                                         & \textbf{0.833}                                            & -                                                   &  -  &  -&  -&  -    \\ \hline
			\multirow{3}{*}{\begin{tabular}[c]{@{}c@{}}57x57\\ (1/4)\end{tabular}}   & psnr & 27.64          & 29.17           & 29.89          & 28.06 & 30.43                                                      & 30.81                                                     & 30.65                                                     & 30.44                                                     & 30.69                                                     & \textbf{30.89}                                           & -& - & - & -   \\ \cline{2-16} 
			& ssim & 0.773          & 0.805           & 0.809          & 0.773 & 0.828                                                      & 0.834                                                     & 0.833                                                     & 0.831                                                     & 0.834                                                     & \textbf{0.837}                                         &   -& - & - & -   \\ \cline{2-16} 
			& vif  & 0.543          & 0.536           & 0.443          & 0.467 & 0.535                                                      & 0.534                                                     & 0.534                                                     & 0.519                                                     & \textbf{0.546}                                            & 0.534       &  -& - & - & -                                                \\ \hline
			\multirow{3}{*}{\begin{tabular}[c]{@{}c@{}}29x29\\ (1/8)\end{tabular}}   & psnr & 24.38          & 25.32           & 26.72 & 24.58 & 25.83                                                      & 26.17                                                     & 26.22                                                     & 26.20                                                     & 26.08                                                     & 26.34     & 25.56 & \textbf{28.31} & - & -  \\ \cline{2-16} 
			& ssim & 0.682          & 0.700           & 0.709          & 0.680 & 0.710                                                      & 0.720                                                     & 0.723                                                     & 0.721                                                     & 0.718                                                     & 0.727                              & 0.707 & \textbf{0.741} & - & -   \\ \cline{2-16} 
			& vif  & \textbf{0.382} & 0.376           & 0.254          & 0.333 & 0.340                                                      & 0.330                                                     & 0.327                                                     & 0.322                                                     & 0.340                                                     & 0.320                               & 0.299& 0.326 & - & -                     \\ \hline
			\multirow{3}{*}{\begin{tabular}[c]{@{}c@{}}15x15\\ (1/16)\end{tabular}}  & psnr & 21.94          & 22.85           & \textbf{24.31} & 22.07 & 23.26                                                      & 20.98                                                     & 23.63                                                     & 23.66                                                     & 23.36                                                     & 20.98                      &   - & - & 22.01 &23.16                              \\ \cline{2-16} 
			& ssim & 0.626          & 0.640           & 0.655          & 0.628 & 0.646                                                      & 0.619                                                     & 0.657                                           & 0.655                                                     & 0.649                                                     & 0.619                                             &  -& - & 0.648 & \textbf{0.670}         \\ \cline{2-16} 
			& vif  & 0.299          & \textbf{0.304} & 0.170          & 0.208 & 0.268                                                      & 0.190                                                     & 0.251                                                     & 0.231                                                     & 0.259                                                     & 0.180                             &       -    & - & 0.218 & 0.260                \\ \hline
		\end{tabular}
	}
\end{table*}

\begin{table*}[]
	\centering
	\caption{Results with different downscaling factors and two different factors for the unwrapped iris region (average values on the test dataset).}
	\label{qualityB}
	\scalebox{0.7}{
		\begin{tabular}{|c|c|c|c|c|c|c|c|c|c|c|c|c|c|c|c|}
			\hline
			\begin{tabular}[c]{@{}c@{}}LR Size\\ (scaling)\end{tabular}              &      & Bilinear & Bicubic        & PCA            & SAE            & \begin{tabular}[c]{@{}c@{}}CNN FS \\ Factor 2\end{tabular} & \begin{tabular}[c]{@{}c@{}}CNN FS\\ Factor 4\end{tabular} & \begin{tabular}[c]{@{}c@{}}CNN TL\\ Factor 2\end{tabular} & \begin{tabular}[c]{@{}c@{}}CNN TL\\ Factor 4\end{tabular} & \begin{tabular}[c]{@{}c@{}}CNN FT\\ Factor 2\end{tabular} & \begin{tabular}[c]{@{}c@{}}CNN FT\\ Factor 4\end{tabular} & \begin{tabular}[c]{@{}c@{}}CNN FS\\ Factor 8\end{tabular} & \begin{tabular}[c]{@{}c@{}}CNN FT\\ Factor 8\end{tabular} & \begin{tabular}[c]{@{}c@{}}CNN FS\\ Factor 16\end{tabular} & \begin{tabular}[c]{@{}c@{}}CNN FT\\ Factor 16\end{tabular} \\ \hline
			\multirow{3}{*}{\begin{tabular}[c]{@{}c@{}}115x115\\ (1/2)\end{tabular}} 
			& psnr & 34.27    & 36.22          & 36.83          & 34.69          & 37.69   & -      & 37.80  & -                                                         & \textbf{38.08}                                            & -    & - & - & - & -                                                      \\ \cline{2-16} 
			& ssim & 0.930    & 0.951          & 0.955          & 0.923          & 0.963   & -      & 0.963  & -                                                         & \textbf{0.964}                                            & -      & - & - & - & -                                                    \\ \cline{2-16} 
			& vif  & 0.812    & 0.848          & 0.824          & 0.766          & 0.859   & -      & 0.864                                                     & -                                                         & \textbf{0.872}                                            & -                                        & - & - & - & -                  \\ \hline
			\multirow{3}{*}{\begin{tabular}[c]{@{}c@{}}57x57\\ (1/4)\end{tabular}}   
			& psnr & 29.27    & 31.14          & 32.13          & 29.94          & 32.76   & 33.34  & 33.02                                                     & 32.73                                                     & 33.03                                                     & \textbf{33.40}       & - & - & - & -                                      \\ \cline{2-16} 
			& ssim & 0.853    & 0.873          & 0.874          & 0.852          & 0.887   & 0.891  & 0.890                                                     & 0.889                                                     & 0.891                                                     & \textbf{0.893}         & - & - & - & -                                    \\ \cline{2-16} 
			& vif  & 0.583    & 0.601          & 0.550          & 0.540          & 0.614   & 0.626  & 0.625                                                     & 0.617                                                     & 0.630                                                     & \textbf{0.632}           & - & - & - & -                                  \\ \hline
			\multirow{3}{*}{\begin{tabular}[c]{@{}c@{}}29x29\\ (1/8)\end{tabular}}   
			& psnr & 25.59    & 26.67          & \textbf{28.61} & 25.86          & 27.25   & 27.56  & 27.74                                                     & 27.73                                                     & 27.56                                                     & 27.91                     & 25.56 & 28.31 & - & -                                 \\ \cline{2-16} 
			& ssim & 0.791    & 0.803          & 0.811          & 0.788          & 0.810   & 0.818  & 0.820                                                     & 0.819                                                     & 0.816                                                     & 0.823                     & 0.806 & \textbf{0.837} & - & -                       \\ \cline{2-16} 
			& vif  & 0.456    & \textbf{0.459} & 0.399          & 0.429          & 0.443   & 0.449  & 0.451                                                     & 0.444                                                     & 0.449                                                     & 0.450                           & 0.425 & 0.440 & - & -                           \\ \hline
			\multirow{3}{*}{\begin{tabular}[c]{@{}c@{}}15x15\\ (1/16)\end{tabular}}  
			& psnr & 22.96    & 23.97          & \textbf{25.82} & 23.08          & 24.42   & 24.55  & 24.94                                                     & 24.96                                                     & 24.55                                                     & 22.15                                    & - & - & 23.31 & 24.67                  \\ \cline{2-16} 
			& ssim & 0.748    & 0.760          & 0.774          & 0.749          & 0.763   & 0.743  & \textbf{0.774}                                            & 0.772                                                     & 0.766                                                     & 0.743                                    & - & - & 0.761 & \textbf{0.785}                  \\ \cline{2-16} 
			& vif  & 0.417    & 0.414          & 0.335          & 0.417   & 0.393                                                   & 0.350                                                     & 0.386                                                     & 0.374                                                     & 0.386                                                     & 0.342                 & - & - & 0.407 & \textbf{0.419}                                    \\ \hline
		\end{tabular}
	}
\end{table*}

For the experiments we use the CASIA Interval v3 iris database that contains a total of 2.655 NIR images of size 280x320 pixels, from 249 subjects captured with a self developed close-up camera, resulting in 396 different eyes. Manual segmentation annotation of the database is available, which is used as input for our experiments.
In the pre-processing step all images are resized via bicubic interpolation in order to have the same sclera radius and are aligned by extracting a square region of 231x231 around the pupil center. All images that do not fit in this requirement (for example when the eye is close to the image border) are discarded. After this, the 1.872 remaining images are used in the experiments. For the deep learning training and tests, the pre-processed dataset is divided into two separated sets: 925 images from the first 116 users for the training and 947 images from the remaining 133 users for the tests (we consider each eye as a different user). This set division by users is important to make sure that the same pattern (in the patches) will not be used both in training and testing steps.

To evaluate the performance of the methods by quality assessment algorithms we use the Peak Signal to Noise Ratio \textbf{(PSNR)} that is the ratio between the peak signal and the power of corrupting noise that affects the fidelity of its representation, the Structural Similarity Index Measure  \textbf{(SSIM)} that extracts three separate scores (visual influence, contrast and structural score) combining them to the final score and the Visual Information Fidelity \textbf{(VIF)} that calculates the mutual information between input and the output of the HVS channel when no distortion is present and the mutual information between the input of the distortion channel and the output of the HVS channel for the test signal \cite{Hofbauer201660}. In these metrics, a high metric score reflects a high quality. For the quality tests, all images from the database were used in high resolution as reference images. We compare our method with bilinear and bicubic interpolation as well as to PCA hallucination of local patches used in \cite{Alonso}.

We also conduct recognition experiments using reconstructed images to evaluate the iris recognition performance. In this procedure, first the iris is unwrapped to a normalized rectangle of 20x240 pixels using the Daugman's rubber sheet model \cite{Daugman}, then a 1D Log-Gabor (LG) wavelet is applied with a phase binary quantization to 4 levels \cite{Masek03recognitionof}.   The comparison between the binary vectors is done by the normalized Hamming Distance \cite{Daugman} where the rotation is accounted for by shifting the grid of the query image in counter- and clock-wise directions, and selecting the lowest distance that corresponds to the best match. We also implemented a SIFT comparator in which SIFT feature points in scale space are extracted from the iris region (without unwrapping) and the comparison is performed based on the texture information around the feature points using the SIFT operator \cite{Alonso2}.

\section{ Results}

\label{sec:Results}
The results of the quality assessment for the test images and for the normalized iris region (20x240) are shown in Table \ref{qualityA}   and Table \ref{qualityB}. It can be seen in Table \ref{qualityA}  that the use of the Convolutional Neural Networks outperforms the traditional methods of interpolation (bicubic and bilinear) as well as the eigen-patch hallucination (PCA) method, mainly for small downscaling factors. It also can be noticed that the use of the Fine Tuning strategy improves the results by merging the use of natural and iris images during the CNN training.
Also, when the CNN is trained with the same downscaling factor as the tests, the results are also becoming more resilient for lower resolutions. It can also be noticed that, for low resolutions, the quality assessment algorithms present different best results which can make the results interpretation difficult.

In iris recognition verification we consider two scenarios: 1) enrollment samples taken from original HR input images, and query samples taken from reconstructed super-resolution results (Table \ref{EER1}) simulating a controlled enrollment scenario (for example, when the user is registered using a HR sensor and make use of the system using a cellphone camera with certain distance); and 2) both enrollment and query samples taken from the reconstructed super-resolution results (Table \ref{EER2}) simulating a totally uncontrolled scenario (for example, when the user is registered using a cellphone and make use of the system also using a cellphone camera with certain distance). 

It can be observed that the performance of CNN's are the best for small downscaling factors in both scenarios in general, despite the diversity of good results among the training approaches. Using the Log-gabor comparator the CNN using Fine Tuning and Transfer Learning approach beats the other methods except for the lowest resolution that PCA does best.
For the SIFT comparator the CNN's are better but there is no particular winning training approach, in this case, using the downscaling factor of 2 the SAE method present the best result for the scenario 2. It also can be seen that for the SIFT comparator the performances of the Bicubic and Bilinear methods degrade rapidly when the resolution decreases, whereas the CNN methods show high resiliency.

It is interesting to notice that in scenario 1 (Table \ref{EER1}), the CNN methods perform better in factor 2 and 4 than using the original images without downscaling which means that it, in terms of recognition, is better to downscale the original image (i.e. apply a blur filter) and apply the deep-learning methods from the sensor before comparison.

\section{Conclusion}

In this work we investigated deep learning single-image super-resolution methods using Stacked Auto-Encoders and Convolutional Neural Networks to increase the resolution of iris images. To address the problem we tested if the end-to-end mapping between low and high resolution images can be successful applied using different strategies as transfer learning and fine-tuning to improve the results. 

Evaluation performed on a database of near-infrared iris images with different upscaling factors both in the training process and in the tests shows the superiority of the tested methods over the compared methods in terms of quality assessment, with the CNN using Fine Tuning approach presenting the best results on average. When we evaluate the recognition rate by iris comparison experiments, the CNN’s in general presented better results but there was no particular CNN approach being the best in all scenarios. We also showed that an uncontrolled scenario (scenario 2 in the EER verification results) is feasible since the deep learning approach in scenario 2 presented better accuracy results than the scenario 1.

Also, it is important to notice that recognition performed is not considerably degraded until image is downscaled by 1/8 or higher factors, allowing to use both query and test images of reduced size which can be an advantage for systems under low storage or data transmission capabilities.

In future work we intend to focus on the Convolutional Neural Network approach trying new methods as the use of recursive layers and investigate the use of other loss functions as perceptual loss functions as well as explore other datasets with different semantic knowledge to perform the fine tuning approach.

% \usepackage{multirow}
\begin{table*}
	\centering
	\caption{Verification results (EER) of the scenario 1 (original vs. downscaled) considered for different downscaling factors. The results for the original database with no scaling for the LG and SIFT are respectively 0.76 and 4.19.}
	\label{EER1}
	\scalebox{0.7}{
		\begin{tabular}{|c|c|c|c|c|c|c|c|c|c|c|c|c|c|c|c|}
			\hline
			\begin{tabular}[c]{@{}c@{}}LR Size\\ (scaling)\end{tabular}               & \multicolumn{1}{l|}{} & Bilinear      & Bicubic       & PCA           & SAE   & \begin{tabular}[c]{@{}c@{}}CNN FS \\ Factor 2\end{tabular} & \begin{tabular}[c]{@{}c@{}}CNN FS\\ Factor 4\end{tabular} & \begin{tabular}[c]{@{}c@{}}CNN TL\\ Factor 2\end{tabular} & \begin{tabular}[c]{@{}c@{}}CNN TL\\ Factor 4\end{tabular} & \begin{tabular}[c]{@{}c@{}}CNN FT\\ Factor 2\end{tabular} & \begin{tabular}[c]{@{}c@{}}CNN FT\\ Factor 4\end{tabular} & \begin{tabular}[c]{@{}c@{}}CNN FS\\ Factor 8\end{tabular} & \begin{tabular}[c]{@{}c@{}}CNN FT\\ Factor 8\end{tabular} & \begin{tabular}[c]{@{}c@{}}CNN FS\\ Factor 16\end{tabular} & \begin{tabular}[c]{@{}c@{}}CNN FT\\ Factor 16\end{tabular} \\ \hline
			%\multirow{2}{*}{NO DS}                                          & LG                 & \multicolumn{14}{c|}{0.76}                                                                                                                                                                                                                                                                                                                                                                                                                                                                                                                                                                                                                                                       \\ \cline{2-16} 
			%& SIFT                  & \multicolumn{14}{c|}{4.19}                                                                                                                                                                                                                                                                                                                                                                                                                                                                                                                                                                                                                                                       \\ \hline
			\multirow{2}{*}{\begin{tabular}[c]{@{}c@{}}115x115 \\ (1/2)\end{tabular}} & LG                 & \textbf{0.69} & \textbf{0.69} & 0.73          & 3.00  & 0.72                                                       & -                                                         & 0.76                                                      & -                                                         & \textbf{0.69}                                             & -                                                         & -                                                         & -                                                         & -                                                          & -                                                          \\ \cline{2-16} 
			& SIFT                  & 4.05          & \textbf{3.51} & 3.81          & 4.21  & 4.01                                                       & -                                                         & 4.21                                                      & -                                                         & 4.01                                                      & -                                                         & -                                                         & -                                                         & -                                                          & -                                                          \\ \hline
			\multirow{2}{*}{\begin{tabular}[c]{@{}c@{}}57x57 \\ (1/4)\end{tabular}}   & LG                 & 0.69          & 0.68          & 0.73          & 1.34  & 0.69                                                       & 0.68                                                      & 0.72                                                      & 0.72                                                      & 0.68                                                      & \textbf{0.67}                                             & -                                                         & -                                                         & -                                                          & -                                                          \\ \cline{2-16} 
			& SIFT                  & 10.42         & 7.41          & 5.20          & 10.13 & 4.95                                                       & \textbf{4.34}                                             & 4.47                                                      & 4.67                                                      & 4.41                                                      & \textbf{4.34}                                             & -                                                         & -                                                         & -                                                          & -                                                          \\ \hline
			\multirow{2}{*}{\begin{tabular}[c]{@{}c@{}}29x29 \\ (1/8)\end{tabular}}   & LG                 & 1.61          & 1.42          & 1.11          & 2.33  & 1.18                                                       & 1.18                                                      & 1.07                                                      & 1.10                                                      & 1.09                                                      & \textbf{1.02}                                             & 1.53                                                      & 1.37                                                      & -                                                          & -                                                          \\ \cline{2-16} 
			& SIFT                  & 28.23         & 24.99         & 15.86         & 35.31 & 17.50                                                      & \textbf{14.26}                                            & 16.31                                                     & 17.34                                                     & 17.96                                                     & 15.87                                                     & 20.65                                                     & 17.65                                                     & -                                                          & -                                                          \\ \hline
			\multirow{2}{*}{\begin{tabular}[c]{@{}c@{}}15x15 \\ (1/16)\end{tabular}}  & LG                 & 10.39         & 9.59          & \textbf{7.29} & 14.29 & 9.07                                                       & 18.72                                                     & 8.96                                                      & 9.67                                                      & 9.43                                                      & 17.84                                                     & -                                                         & -                                                         & 19.53                                                      & 15.74                                                      \\ \cline{2-16} 
			& SIFT                  & 50.52         & 47.33         & 36.51         & 48.02 & 41.76                                                      & 42.06                                                     & 38.23                                                     & \textbf{36.36}                                            & 40.99                                                     & 39.08                                                     & -                                                         & -                                                         & 42.60                                                      & 45.35                                                      \\ \hline
		\end{tabular}
	}
\end{table*}

% \usepackage{multirow}
\begin{table*}
	\centering
	\caption{Verification results (EER) of the scenario 2 (downscaled vs. downscaled) considered for different downscaling factors.  The results for the original database with no scaling for the LG and SIFT are respectively 0.76 and 4.19.}
	\label{EER2}
	\scalebox{0.7}{
		\begin{tabular}{|c|c|c|c|c|c|c|c|c|c|c|c|c|c|c|c|}
			\hline
			\begin{tabular}[c]{@{}c@{}}LR Size\\ (scaling)\end{tabular}               & \multicolumn{1}{l|}{} & Bilinear      & Bicubic & PCA           & SAE           & \begin{tabular}[c]{@{}c@{}}CNN FS \\ Factor 2\end{tabular} & \begin{tabular}[c]{@{}c@{}}CNN FS\\ Factor 4\end{tabular} & \begin{tabular}[c]{@{}c@{}}CNN TL\\ Factor 2\end{tabular} & \begin{tabular}[c]{@{}c@{}}CNN TL\\ Factor 4\end{tabular} & \begin{tabular}[c]{@{}c@{}}CNN FT\\ Factor 2\end{tabular} & \begin{tabular}[c]{@{}c@{}}CNN FT\\ Factor 4\end{tabular} & \begin{tabular}[c]{@{}c@{}}CNN FS\\ Factor 8\end{tabular} & \begin{tabular}[c]{@{}c@{}}CNN FT\\ Factor 8\end{tabular} & \begin{tabular}[c]{@{}c@{}}CNN FS\\ Factor 16\end{tabular} & \begin{tabular}[c]{@{}c@{}}CNN FT\\ Factor 16\end{tabular} \\ \hline
		%	\multirow{2}{*}{NO DS}                                          & LG                 & \multicolumn{14}{c|}{0.76}                                                                                                                                                                                                                                                                                                                                                                                                                                                                                                                                                                                                                                                         \\ \cline{2-16} 
		%	& SIFT                  & \multicolumn{14}{c|}{4.19}                                                                                                                                                                                                                                                                                                                                                                                                                                                                                                                                                                                                                                                         \\ \hline
			\multirow{2}{*}{\begin{tabular}[c]{@{}c@{}}115x115 \\ (1/2)\end{tabular}} & LG                 & \textbf{0.61} & 0.73    & 0.72          & 0.66          & 0.72                                                       & -                                                         & 0.72                                                      & -                                                         & 0.72                                                      & -                                                         & -                                                         & -                                                         & -                                                          & -                                                          \\ \cline{2-16} 
			& SIFT                  & 3.01          & 3.13    & 3.71          & \textbf{2.54} & 3.82                                                       & -                                                         & 3.80                                                      & -                                                         & 3.82                                                      & -                                                         & -                                                         & -                                                         & -                                                          & -                                                          \\ \hline
			\multirow{2}{*}{\begin{tabular}[c]{@{}c@{}}57x57 \\ (1/4)\end{tabular}}   & LG                 & 0.76          & 0.65    & 0.68          & 0.72          & 0.68                                                       & 0.65                                                      & \textbf{0.60}                                             & 0.66                                                      & 0.62                                                      & 0.68                                                      & -                                                         & -                                                         & -                                                          & -                                                          \\ \cline{2-16} 
			& SIFT                  & 4.26          & 3.08    & 3.37          & 3.45          & 2.09                                                       & 2.23                                                      & 2.41                                                      & 2.29                                                      & \textbf{1.94}                                             & 2.50                                                      & -                                                         & -                                                         & -                                                          & -                                                          \\ \hline
			\multirow{2}{*}{\begin{tabular}[c]{@{}c@{}}29x29 \\ (1/8)\end{tabular}}   & LG                 & 2.38          & 1.88    & 1.18          & 2.14          & 1.30                                                       & 1.95                                                      & \textbf{0.98}                                             & 1.24                                                      & 1.14                                                      & 1.26                                                      & 1.71                                                      & 1.41                                                      & -                                                          & -                                                          \\ \cline{2-16} 
			& SIFT                  & 14.82         & 11.6    & 7.54          & 15.82         & 6.50                                                       & \textbf{6.26}                                             & 7.33                                                      & 8.14                                                      & 7.30                                                      & 7.26                                                      & 8.65                                                      & 7.45                                                      & -                                                          & -                                                          \\ \hline
			\multirow{2}{*}{\begin{tabular}[c]{@{}c@{}}15x15 \\ (1/16)\end{tabular}}  & LG                 & 11.03         & 11.25   & \textbf{4.79} & 8.58          & 9.10                                                       & 14.31                                                     & 6.26                                                      & 8.18                                                      & 7.88                                                      & 11.64                                                     & -                                                         & -                                                         & 12.43                                                      & 11.46                                                      \\ \cline{2-16} 
			& SIFT                  & 41.66         & 36.37   & 19.50         & 36.35         & 22.64                                                      & 20.12                                                     & 22.28                                                     & \textbf{17.26}                                            & 22.78                                                     & 19.08                                                     & -                                                         & -                                                         & 19.59                                                      & 26.40                                                      \\ \hline
			
		\end{tabular}
	}
\end{table*}

\section*{Acknowledgment}
This research was partially supported by CNPq-Brazil for Eduardo Ribeiro under grant No. 00736/2014-0.

\bibliographystyle{IEEEbib}
\bibliography{refs}

\end{document}